\documentclass[
aps,
prl,
twocolumn,
superscriptaddress,
nofootinbib,
nobibnotes,
longbibliography,
showkeys
]{revtex4-2}

\usepackage[utf8]{inputenc} 
\usepackage[T1]{fontenc}    
\usepackage[dvipsnames]{xcolor}   
\usepackage[
    colorlinks=true,        
    linkcolor=red,
    citecolor=blue,
    urlcolor=BrickRed
]{hyperref}
\usepackage{url}            
\usepackage{booktabs}       
\usepackage{amsfonts}       
\usepackage{nicefrac}       
\usepackage{microtype}      
\usepackage{amsmath}
\usepackage{graphicx} 
\usepackage{caption}  
\usepackage{float}    
\usepackage{subcaption} 
\usepackage{booktabs} 
\usepackage{enumitem}

\makeatletter
\let\newfloat\newfloat@ltx
\makeatother

\usepackage{algorithm}      
\usepackage{algpseudocode}  
\usepackage{algorithmicx}   

\usepackage{rotating}
\usepackage{tabularx}
\usepackage{booktabs}
\usepackage{adjustbox}
\usepackage{multirow}
\usepackage{dsfont}

\makeatletter

\def\@affilcomma#1#2{%
  \@ifx{\z@#1}{%
    \@ifx{\relax#2}{}{%
      \@affilcomma{#2}%
    }%
  }{%
    \@alph{#1}%
    \@ifx{\relax#2}{}{%
      \@ifx{\z@#2}{%
        \@affilcomma
      }{%
        ,\,\@affilcomma{#2}%
      }%
    }%
  }%
}

\patchcmd{\@affil@script}
  {\def\@thefnmark{#1}}
  {\def\@thefnmark{\@alph{#1}}}
  {}{}

\makeatother

\begin{document}

\title{Detecting Early Warning Signals by Tracing Supply Chain Disruptions from Minimally Shared Information using a Multi-Agent Framework}

\author{Guangya Hao}
\affiliation{Department of Engineering, University of Cambridge, Cambridge, UK}
\author{Yunbo Long}
\affiliation{Department of Engineering, University of Cambridge, Cambridge, UK}
\author{Ge Zheng}
\affiliation{Department of Engineering, University of Cambridge, Cambridge, UK}
\author{Tianxing Ma}
\affiliation{Department of Engineering, University of Cambridge, Cambridge, UK}
\author{Alexandra Brintrup}
\email[Corresponding author: ]{ab702@cam.ac.uk}
\affiliation{Department of Engineering, University of Cambridge, Cambridge, UK}
\affiliation{The Alan Turing Institute, London, UK}

\begin{abstract}
Modern supply chains are decentralised, privacy-sensitive networks in which disruptions on individual supplier--buyer edges can propagate downstream and cause significant operational losses. 
Early detection of disruptions is difficult because no single actor has global visibility of the network, and raw operational data are rarely shared across firms. 
To address this challenge, we propose an early-warning framework for disruption detection and localisation 
under limited information sharing between multiple agents. 
In this framework, each firm converts its local inventory time series into a lightweight degree-distribution summary using a recurrence network. 
Firms in the same tier exchange only privacy-preserving summaries to detect significant increases in cross-firm dynamical similarity  
via changepoint detection. 
Once a warning signal is triggered, the framework clusters impacted firms and traces evidence upstream to localise the disrupted edge.
To evaluate the method, we develop an agent-based simulator and construct a dynamic inventory dataset based on four-tier supply chain instances. 
Experimental results show that the proposed framework can effectively localise disrupted edges under both single-edge and multi-edge disruption settings, and outperform a local inventory-threshold baseline. 
Further robustness studies show that detection performance is moderated by disruption severity, network topology, and the distance between the downstream firm and the disrupted edge. In particular, cross-tier connectivity reduces detection performance and increases localisation delays, while high detection rates are maintained under scale-free topology. Future work will address the current limitations of synthetic data and acyclic structures by incorporating real-world operational data and more complex network topologies.

\end{abstract}


\maketitle

\section{Introduction}\label{sec:introduction}
Modern supply chains are exposed to localised disruptions such as supplier shutdowns, transport delays, and capacity reductions. 
Although most disruptions often occur on a single supplier–buyer relationship, even disruptions on a single edge can propagate through multiple downstream firms due to delayed inventory shortages and synchronised stress~\citep{elliott2022supply,hao2026continual}. 
Therefore, early warning systems should not only detect that a disruption has occurred, but also localise the disrupted supplier–buyer edge \cite{Ivanov2014,IvanovDolgui2020}.

Undertaking such a task is challenging because supply-chain disruptions are temporal and decentralised.
A disruption on a supplier--buyer relationship may propagate through the network with delay, and its early impact can be partially absorbed by buffer inventory, alternative supply paths, or demand fluctuations \cite{Ivanov2014, Hosseini2020, Ivanov2021}.
As a result, the disruption may only later appear as a change in the temporal pattern of local operations.
Therefore, early warning requires monitoring evolving temporal dynamics rather than only checking whether a single inventory value crosses an abnormal threshold \cite{Ashraf2024, Alqudah2025}.
Moreover, supply-chain data are distributed across firms and are commercially sensitive.
Firms usually observe only their own operations and are often unwilling or unable to share raw inventory trajectories, order quantities, shipment records, or detailed supplier--buyer relationships with a central platform~\cite{Zheng2023FederatedRisk,Zheng2025AdaptiveFL,Zheng2025GraphVisibility}.

Existing approaches to propagation modelling and early warning systems do not fully resolve these constraints. 
centralised digital-twin and graph-learning methods can model disruption propagation, but they often assume global visibility or aggregated cross-firm data \cite{IvanovDolgui2020,Ashraf2024,Kosasih2022,Alqudah2025}.
Federated-learning approaches mitigate the need for direct data sharing, but they mainly focus on collaborative risk prediction or model training rather than event-driven disruption monitoring \cite{Zheng2023FederatedRisk,Kong2024SCF, Zheng2025AdaptiveFL, Zheng2025GraphVisibility}.
Early-warning methods based on temporal synchronisation and recurrence structures provide useful indicators, however they are usually designed for settings where all relevant time series can be jointly analysed~\cite{Donner2010RecurrenceNetworks, Gatfaoui2019Flickering, Song2024MRNEWS}.
These bottlenecks raise the following central question:

\textit{How can supply-chain disruptions be detected early and localised in a distributed manner when firms only share limited local information rather than full operational data?}

To address this question, this paper proposes a privacy-preserving multi-agent early-warning framework in which each node monitors only its own local inventory time series and communicates only lightweight summaries with a limited number of network neighbours. 
Specifically, each node transforms its temporal observations into a recurrence-network representation, exchanges compressed topological summaries with same-tier neighbours, and measures cross-firm dynamical similarity by comparing the shared degree distributions.
%
A changepoint in tier-level similarity serves as an early-warning trigger, after which local spectral clustering and hierarchical upstream reporting are used to progressively localise the disrupted edge. 
In this way, the framework avoids centralised coordination and raw-data sharing whilst still enabling timely disruption detection and interpretable localisation.
Consider a silicon supply chain in which silicon materials are processed into wafers, wafers are supplied to chip manufacturers, and the resulting chips are delivered to downstream manufacturers. Suppose that a supplier–buyer edge $e^\star = (u^\star, v^\star)$, connecting a wafer manufacturer to a chip manufacturer, is disrupted. Immediately after the disruption, its local operational effect may be partially masked by safety stock, in-transit inventory, or alternative supply arrangements. However, as the resulting shortage propagates through the network, multiple downstream firms depending on the same supply path may begin to exhibit correlated changes in inventory dynamics.  A common upstream disruption may produce only a weak anomaly at each individual firm, but it can create a broader and more synchronised operational footprint across multiple downstream firms. The framework therefore uses this downstream coordination pattern as the initial warning evidence. 

Under the proposed framework, these firms do not exchange raw inventory trajectories, order quantities, or shipment records. Instead, each firm converts its local time series into a compressed recurrence-network summary and shares only this summary with permitted same-tier peers. Abnormal increases in cross-firm dynamic similarity are used to detect a common upstream problem, after which the affected-firm evidence is traced upstream to localise the disrupted supplier–buyer relationship. No firm is assumed to know the disrupted edge in advance. When the method localises $e^\star$, the precise edge-level result can be disclosed only to the firms incident to that relationship, while more distant downstream firms receive only an alert that an upstream abnormality has been detected. The example therefore illustrates two forms of limited information sharing: firms exchange compressed operational summaries rather than raw data, and the final localisation result is disclosed at different levels of detail according to each firm’s role.
Experimental results demonstrate that the proposed framework can effectively localise disrupted supplier–buyer edges under both single-edge and multi-edge disruption settings, outperforming a local inventory-threshold baseline. Further robustness studies show that higher disruption severity generally leads to shorter localisation delays, while cross-tier connectivity reduces detection rates and increases localisation delays compared with strictly hierarchical topology. In contrast, the framework maintains a high detection rate under scale-free topology, whose characteristics have been observed in real-world supply networks. The results also show that detection rates and localisation delays do not vary monotonically with the distance from the disrupted edge.
The main contributions of this work are summarised as follows:
\begin{itemize}
    \item %
    We propose a distributed monitoring framework that preserves commercial privacy by avoiding raw operational data sharing, while also formulating the problem of early edge-disruption detection and localisation in decentralised supply networks.
    \item %
    We introduce a temporal-topological detection methodology that combines local recurrence-network modelling, privacy-preserving degree-distribution similarity estimation, changepoint-based triggering, local spectral extraction of correlated firms, and hierarchical edge localisation.
    This design enables both early warning and interpretable localisation under limited communication.
    \item %
    We develop an open-source agent-based simulation framework that dynamically models inventory evolution under supply disruptions, and we construct a dynamic inventory benchmark dataset to support reproducible evaluation of distributed disruption monitoring and localisation methods.
\end{itemize}

The remainder of this paper is organised as follows. Section~\ref{sec:relatedwork} reviews the related work and highlights the current research gaps. Section~\ref{sec:model} formally defines the system model, problem statement, and the foundational concepts of recurrence networks used in our approach. Section~\ref{sec:methods} details the proposed multi-agent distributed monitoring architecture and methodology. 
Section~\ref{sec:setup} presents the experimental setup, while Section~\ref{sec:results} reports the experimental results. Finally, Section~\ref{sec:conclusion} concludes the paper with discussions on limitations and future research directions.

\section{Related Work}
\label{sec:relatedwork}

\subsection{Centralised Supply Chain Disruption Detection}
Many studies have examined how supply-chain disruptions are detected and how their effects propagate through supply networks\citep{hendricks2005empirical,hendricks2005association}.
Classical ripple-effect studies typically use simulation, Bayesian networks, or dynamic Bayesian networks to analyse how disruptions spread across supply structures \cite{Hosseini2020RippleDBN}.
More recent machine-learning approaches extended this line by using firm-level historical records to predict supplier disruptions \cite{Brintrup2020SupplierDisruptions}, embedding disruption monitoring into cognitive digital supply chain twins \cite{Ashraf2024CDSCT}, or applying graph neural networks to estimate post-hazard firm-level outcomes from relational and exogenous information \cite{Yang2024GNNHazard}.
These studies show that network structure and temporal operational signals are important for disruption assessment\citep{artime2024robustness, elliott2022supply}.
However, they generally rely on centralised data access, global network visibility, or a platform capable of aggregating cross-firm information.
Such assumptions are difficult to satisfy in real supply-chain ecosystems, where operational data and supply relationships are distributed across firms.
As a consequence, even strong centralised predictors may be difficult to deploy operationally.
In contrast, our work focuses on distributed early warning and disruption localisation under limited information sharing.

\subsection{Privacy-Preserving Collaboration in Supply Chain Analytics}

Federated learning has recently been used to support collaborative supply-chain analytics without requiring firms to share raw operational data.
Prior studies have applied this idea to collective supply-chain risk prediction \cite{Zheng2023FederatedRisk}, order-level risk prediction in supply-chain financing \cite{Kong2024SCF}, adaptive information sharing \cite{Zheng2025AdaptiveFL,hao2026self}, and graph-based supply-chain visibility enhancement \cite{Zheng2025GraphVisibility}.
These works demonstrate the feasibility of privacy-conscious collaboration for risk preparedness in supply-chain ecosystems.
Nevertheless, existing privacy-preserving methods mainly focus on prediction, visibility enhancement, or model training.
They are not primarily designed for event-driven disruption monitoring, where the goal is to detect an emerging disruption and localise its source under limited communication.
Moreover, many federated approaches still require a central aggregator or repeated training rounds.
Our framework instead avoids both raw-data exchange and centralised parameter aggregation, because firms share only lightweight recurrence-network summaries, detect abnormal peer similarity, and propagate local evidence upstream to localise disrupted supplier--buyer edges.

\subsection{Early-Warning Signals in Complex Systems}

Early-warning signals for critical transitions have been widely studied in complex systems, ecology, and finance.
traditional approaches focus on critical slowing down near tipping points, using indicators such as rising autocorrelation, variance, or recovery time \cite{scheffer2009early,kuehn2011critical,dakos2012methods}.
These methods generally assume that a relevant system-level variable is observable and evolves near a slowly changing equilibrium.
However, this assumption is less suitable for decentralised supply chains, where disruptions may emerge through nonlinear, delayed, and partially observed changes across multiple firms.
Network-based approaches offer a broader alternative by mapping temporal dynamics into changes in topology.
Recurrence networks have been used to capture nonlinear time-series behaviour \cite{Marwan2007}, while deviations from quasi-stationary graph structure, topological flickering , and multiplex recurrence indicators have been shown to provide early warnings of financial instability and market crashes \cite{squartini2015stationarity,Gatfaoui2019Flickering,Song2024}.
These studies suggest that rising similarity, synchronisation, or topological instability can reveal systemic stress before a full transition occurs.
More recently, AlMahri et al.~\cite{almahri2026automating} explored agentic AI for supply-chain disruption surveillance using publicly available information. However, such approaches depend on disruptions becoming publicly observable, which is not always guaranteed, particularly during their early stages.
Meanwhile,, existing early-warning methods are usually developed for settings where relevant time series can be collected and jointly analysed.
This assumption does not directly hold in supply chains, where operational data are distributed across firms and raw trajectories cannot normally be centralised.
We therefore adapt temporal-topological early-warning methods to a decentralised setting: each firm constructs a recurrence-network summary from local data, shares only compressed degree-distribution information, and combines changepoint detection \cite{Killick2012PELT,Haynes2017CROPS} with local spectral partitioning \cite{Luxburg2007Spectral} to identify affected firms and trace disruption evidence upstream. 
%

In summary, prior work provides three useful but separate directions: centralised disruption modelling, privacy-preserving collaborative prediction, and early-warning analysis in complex systems.
centralised methods can model disruption propagation, but typically require global visibility or aggregated cross-firm data.
Privacy-preserving learning reduces raw-data sharing, but is mainly designed for collaborative prediction or model training rather than event-driven monitoring.
Early-warning methods reveal temporal synchronisation and topological instability, but usually assume that all relevant time series can be analysed in one place.
As a result, it remains unclear how to detect and localise supply-chain disruptions in a distributed setting where firms share only limited local summaries.
We address this gap with a multi-agent framework that combines local recurrence-based summaries, peer similarity monitoring, and hierarchical upstream evidence tracing.

\section{Problem Formulation and Model}\label{sec:model}

This section formalises the distributed edge-disruption detection problem studied in this paper. 

The formulation is organised into three components: we first introduce the supply network model and define persistent edge-level disruptions (\ref{preliminary}). We then specify the decentralised information structure under which firms observe only local trajectories and communicate privacy-preserving summaries (\ref{distributed information structure}). Based on these two components, we finally formulate disruption monitoring as a distributed detection and localisation problem that balances detection delay, localisation accuracy, and privacy constraints (\ref{distributed detection and localisation}).

\subsection{Preliminary}~\label{preliminary}
We model the supply chain as a directed acyclic graph (DAG)
\begin{equation*}
G=(V,E),
\end{equation*}
where $V=\{1,\dots,n\}$ is the set of firms and $E\subseteq V\times V$ is the set of directed supplier--buyer relationships. A directed edge $e=(u,v)\in E$ indicates that node $u$ supplies material, components, or information to node $v$. 
For each node $v, u\in V$, let
\begin{equation*}
\begin{aligned}
\operatorname{pa}(v)&=\{u\in V:(u,v)\in E\},\\
\operatorname{ch}(v)&=\{w\in V:(v,w)\in E\}
\end{aligned}
\end{equation*}
denote its upstream and downstream neighbourhoods, respectively. 
We further define a tier index $\ell(v)\in \mathbb{N}$ as the graph distance from source nodes, and denote the same-tier set of node $v$ by
\begin{equation*}
\mathcal{S}(v)=\{u\in V:\ell(u)=\ell(v),\ u\neq v\}.
\end{equation*}
Each node $v$ generates a discrete-time local operational signal
\begin{equation*}
x_v(t)\in\mathbb{R}^d,\qquad t=1,2,\dots,T,
\end{equation*}
where $x_v(t)$ may represent inventory level, replenishment state, order-arrival rate, or a low-dimensional vector of locally observable operational variables. 
The local dynamics at node $v$ are modelled abstractly as
\begin{equation*}
x_v(t+1)=F_v\!\Big(x_v(t),\,\{q_{uv}(t)\}_{u\in \operatorname{pa}(v)},\,\xi_v(t)\Big),
\end{equation*}
where $F_v$ denotes the potentially nonlinear and unknown
node-specific transition function, $q_{uv}(t)$ is the effective inflow on edge $(u,v)$, and $\xi_v(t)$ denotes local exogenous disturbances such as demand fluctuations or process noise. 
This formulation intentionally avoids assuming a fully known parametric model, since in practice each firm only observes its own local trajectory.
We further define an edge disruption as a persistent reduction or interruption of the effective transaction flow on a directed supplier–buyer edge. 
For each edge $e=(u,v)\in E$, define a binary disruption state
\begin{equation*}
z_e(t)\in\{0,1\},
\end{equation*}
where $z_e(t)=1$ indicates that edge $e$ is disrupted at time $t$, and $z_e(t)=0$ otherwise. The effective flow on edge $e=(u,v)$ is then written as
\begin{equation*}
q_{uv}(t)=\bar q_{uv}(t)\bigl(1-z_{(u,v)}(t)\bigr),
\end{equation*}
where $\bar q_{uv}(t)$ is the nominal flow in the absence of disruption.
Let
\begin{equation*}
\mathcal{D}^\star(t)=\{e\in E:z_e(t)=1\}
\end{equation*}
denote the true disrupted-edge set at time $t$. In the single-disruption setting,
\begin{equation*}
\mathcal{D}^\star(t)=\{e^\star\},\qquad t\ge t_0,
\end{equation*}
for some unknown disrupted edge $e^\star\in E$ and unknown onset time $t_0$. 
More generally, in the multi-disruption setting, $\mathcal{D}^\star(t)$ is a sparse subset of $E$. 
We focus on persistent disruptions over the detection window, namely
\begin{equation*}
z_e(t)=1,\quad \forall t\ge t_e
\end{equation*}
for each disrupted edge $e$, where $t_e$ is its onset time.
A key difficulty is that an edge disruption may not induce an immediate anomaly at its upstream endpoint. 
Instead, its effect is typically manifested through delayed changes in downstream node trajectories.
Therefore, the disrupted edge is only indirectly observable through the joint temporal behaviour of multiple nodes.

\subsection{Distributed Information Structure}~\label{distributed information structure}
In this work, sensitive data refers to firm-level operational information that may disclose a firm's commercial status, production capacity, supplier dependence, or customer demand.
Such information includes raw inventory trajectories, order quantities, shipment records, replenishment states, demand signals, and detailed supplier--buyer relationships.
To avoid exposing these data, the proposed framework does not exchange raw operational trajectories. Instead, firms communicate only compressed topological summaries extracted from local recurrence networks.
Here, limited shared information has two dimensions: firms exchange compressed summaries rather than raw operational data, and disruption information is disclosed at different levels of granularity. In the silicon supply-chain example, downstream firms use recurrence-network summaries to detect a common upstream abnormality, while hierarchical tracing localises the disrupted edge. Edge-specific information is provided only to the firms incident to that relationship, whereas other downstream firms receive only a general disruption alert.
The monitoring architecture is decentralised. 
There is no central coordinator with access to the full graph or all node-level observations. 
At time $t$, node $v$ has access only to:
\begin{enumerate}
    \item its own observation history
    \begin{equation*}
    x_v(1:t)=\{x_v(1),x_v(2),\dots,x_v(t)\},
    \end{equation*}

    \item its local structural neighbourhood $\operatorname{pa}(v)$, $\operatorname{ch}(v)$, and same-tier communication peers in $\mathcal{S}(v)$,

    \item messages received from its communication neighbourhood.
\end{enumerate}
Let $\mathcal{N}_v\subseteq \operatorname{pa}(v)\cup \operatorname{ch}(v)\cup \mathcal{S}(v)$ denote the nodes with which $v$ is allowed to communicate. 
At each time $t$, node $v$ sends a message
\begin{equation*}
m_{v\to u}(t)=\phi_{v\to u}\bigl(\mathcal{H}_v(t)\bigr),\qquad u\in\mathcal{N}_v,
\end{equation*}
where
\begin{equation*}
\mathcal{H}_v(t)=\Bigl(x_v(1:t),\,\{m_{w\to v}(1:t-1)\}_{w\in\mathcal{N}_v}\Bigr)
\end{equation*}
is the local information history available to node $v$.
The privacy constraint requires that raw operational trajectories are not directly shared. Formally, for all $u\in\mathcal{N}_v$,
\begin{equation*}
m_{v\to u}(t)\neq x_v(1:t),
\end{equation*}
and $m_{v\to u}(t)$ must be a compressed representation derived from local observations. 
In our method, this compressed representation is a topological summary extracted from a recurrence network.

\subsection{Distributed Detection and localisation Problem}~\label{distributed detection and localisation}
Each node $v$ implements a local policy $\pi_v$ that maps its information history $\mathcal{H}_v(t)$ to:
\begin{enumerate}
    \item an outgoing message set $\{m_{v\to u}(t)\}_{u\in\mathcal{N}_v}$,
    \item a local alert variable $a_v(t)\in\{0,1\}$,
    \item a local candidate disrupted-edge set $\widehat{\mathcal{D}}_v(t)\subseteq E$.
\end{enumerate}
A distributed policy is then
\begin{equation*}
\pi=\{\pi_v\}_{v\in V}.
\end{equation*}
Aggregating all local outputs yields a system-level candidate set
\begin{equation*}
\widehat{\mathcal{D}}(t)=\bigcup_{v\in V}\widehat{\mathcal{D}}_v(t).
\end{equation*}
The goal is to design a distributed policy $\pi$ such that the system:
\begin{enumerate}
    \item detects the onset of disruption as early as possible,
    \item localises the true disrupted-edge set as accurately as possible,
    \item minimises communication overhead,
    \item satisfies privacy-preserving information-sharing constraints.
\end{enumerate}
Let $\tau(\pi)$ denote the stopping time at which the system first raises a disruption alert, and let $\widehat{\mathcal{D}}_{\tau}$ be the corresponding estimated disrupted-edge set.
We define the detection delay as
\begin{equation*}
\Delta(\pi)=\bigl(\tau(\pi)-t_0\bigr)^+,
\end{equation*}
where $(x)^+=\max\{x,0\}$.
For localisation error, a natural set-based loss is
\begin{equation*}
L_{\mathrm{loc}}(\widehat{\mathcal{D}}_{\tau},\mathcal{D}^\star)
=\frac{\left|\widehat{\mathcal{D}}_{\tau}\,\triangle\,\mathcal{D}^\star\right|}
{\left|\widehat{\mathcal{D}}_{\tau}\,\cup\,\mathcal{D}^\star\right|+\varepsilon}.
\end{equation*}
where $\triangle$ denotes symmetric difference and $\varepsilon>0$ avoids division by zero.
In the single-edge case, this reduces to the indicator loss
\begin{equation*}
L_{\mathrm{loc}}(\widehat e_{\tau},e^\star)=\mathbf{1}\{\widehat e_{\tau}\neq e^\star\}.
\end{equation*}
Let the cumulative communication cost under policy $\pi$ up to time $\tau$ be
\begin{equation*}
M(\pi)=\sum_{t=1}^{\tau}\sum_{v\in V}\sum_{u\in\mathcal{N}_v}
 c\!\left(m_{v\to u}(t)\right).
\end{equation*}
where $c(\cdot)$ measures message size, transmission count, or a weighted communication burden.
The distributed edge-disruption monitoring problem can therefore be formulated as
\begin{equation*}
\begin{aligned}
\min_{\pi}\; \mathbb{E}\!\Big[&\alpha\,\Delta(\pi)
+\beta\,L_{\mathrm{loc}}(\widehat{\mathcal{D}}_{\tau},\mathcal{D}^\star)\\
&+\gamma\,M(\pi)\Big]
\end{aligned}
\end{equation*}
subject to
\begin{equation*}
m_{v\to u}(t)\in \mathcal{M}_{v\to u},\qquad
m_{v\to u}(t)\ \text{is privacy-preserving},
\end{equation*}
for all $v\in V$, $u\in\mathcal{N}_v$, and $t\le \tau$, where $\alpha,\beta,\gamma>0$ control the trade-off between timeliness, localisation accuracy, and communication cost.
To explicitly discourage false alarms, the policy is further required to satisfy a healthy-network constraint of the form
\begin{equation*}
\mathbb{E}_{\emptyset}[\tau(\pi)]\ge \kappa,
\end{equation*}
where $\mathbb{E}_{\emptyset}$ denotes expectation under the no-disruption regime and $\kappa$ is a prescribed lower bound on mean time to false alarm.
\paragraph{modelling Assumptions}
The above formulation is developed under the following assumptions:
\begin{enumerate}
    \item \textbf{Acyclic topology:} The supply network can be approximated as a DAG over the monitoring horizon.

    \item \textbf{Local observability:} Each node observes at least one local operational time series sampled at regular intervals.

    \item \textbf{Communication locality:} Nodes communicate only with direct upstream/downstream partners and selected same-tier peers.

    \item \textbf{Sparse disruptions:} The number of simultaneously disrupted edges is small relative to $|E|$.

    \item \textbf{Finite-lag detectability:} For each disrupted edge $e\in\mathcal{D}^\star$, there exists a finite lag $\delta_e$ and at least one downstream-affected node set $\mathcal{A}(e)\subseteq V$ such that the distribution of local observations changes after disruption, i.e.,
    \begin{equation*}
    \begin{aligned}
    \mathbb{P}\!\left(x_v(t:t+L)\in \cdot \,\middle|\, e\in\mathcal{D}^\star\right)
    &\neq \\
    \mathbb{P}\!\left(x_v(t:t+L)\in \cdot \,\middle|\, e\notin\mathcal{D}^\star\right).
    \end{aligned}
    \end{equation*}
    for $v\in\mathcal{A}(e)$ and $t\ge t_e+\delta_e$.
\end{enumerate}
These assumptions define the intended deployment regime of the method and clarify that the problem is not one of centralised anomaly classification, but of distributed inference over partially observable, privacy-constrained, and dynamically coupled supply networks.

\section{Methodology}\label{sec:methods}

\subsection{Overview}
\begin{figure*}[t]
\centering
\includegraphics[width=1\textwidth]{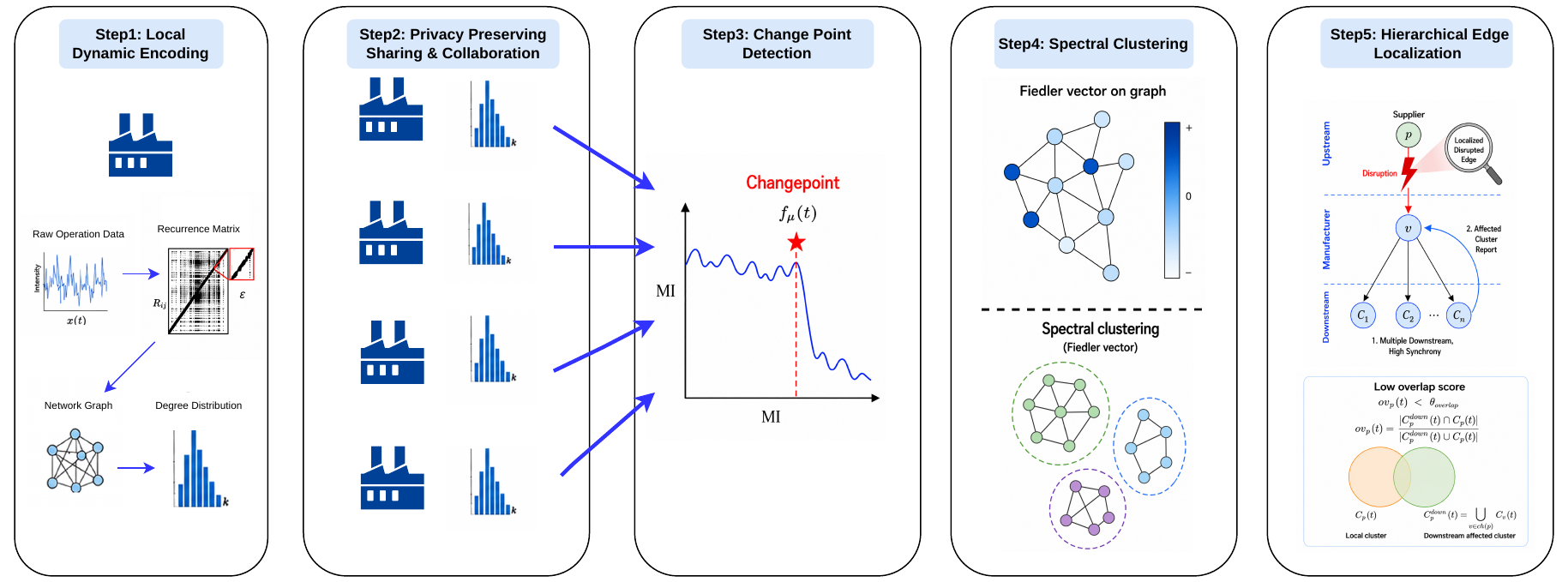}
\caption{Overview of the distributed supply-chain monitoring framework used in the methodology section. Each firm monitors its local inventory dynamics, exchanges lightweight summaries with limited peers, and participates in decentralised disruption detection and localisation.}
\label{fig:method_overview}
\end{figure*}

We propose a distributed early-warning framework for detecting and localising edge disruptions in privacy-constrained supply networks. 
Each node uses only its own local inventory trajectory and exchanges only lightweight summaries with a limited set of same-tier or neighboring nodes. 
The method consists of five steps: \textit{Step 1}, local recurrence-network construction, \textit{Step 2}, privacy preserving sharing , \textit{Step 3}, changepoint detection, \textit{Step 4}, local extraction of affected clusters and \textit{Step 5}, hierarchical edge localisation. 
The workflow is event-driven: under normal conditions, nodes communicate only compact summaries; once an abnormal coordination pattern is detected, they switch to localisation mode. 
This design keeps the framework decentralised, privacy-preserving, and communication-efficient. An overview of the methodology is shown in Figure~\ref{fig:method_overview}.
Algorithm ~\ref{alg:node_level} details the monitoring and localisation procedure executed by each node, while Algorithm ~\ref{alg:system_level} describes how node-level alerts and candidate-edge reports are aggregated across the network.
\paragraph{\textbf{Step 1: Local Dynamic Encoding}}
The first step is to convert each node's local inventory trajectory into a compact representation of its recent dynamics. For node $v$, at decision time $t$, we consider a sliding window
\begin{equation*}
W_v(t)=\{x_v(t-L+1),\ldots,x_v(t)\},
\end{equation*}
where $L$ is the window length. This window contains the recent inventory evolution of node $v$.
To capture temporal dynamics rather than only raw values, the windowed series is mapped into phase space using recurrence analysis 
\cite{Donner2010RecurrenceNetworks}. 
More generally, delay-coordinate embedding is adopted:
\begin{equation*}
\begin{aligned}
\mathbf{y}_i={}&\bigl[x_v(t_i),\,x_v(t_i+\tau),\,\ldots,\\
&x_v(t_i+(m-1)\tau)\bigr],\qquad i=1,\ldots,N.
\end{aligned}
\end{equation*}
where $m$ is the embedding dimension, $\tau$ is the delay, and
\begin{equation*}
N=L-(m-1)\tau.
\end{equation*}
Each vector $\mathbf{y}_i$ represents the local dynamical state of node $v$ at one time point.
A recurrence matrix is then constructed as
\begin{equation*}
R_{ij}^{(v)}(t)=\Theta\!\bigl(\epsilon_v(t)-\|\mathbf{y}_i-\mathbf{y}_j\|\bigr),
\qquad i\neq j,
\end{equation*}
where $\Theta(\cdot)$ is the heaviside step function and $\epsilon_v(t)$ is a threshold selected to satisfy a target recurrence rate $\rho$.
Thus, two time points are connected if their local states are sufficiently similar. 
The recurrence matrix defines a recurrence network whose nodes correspond to embedded time points and whose edges indicate recurrent visits to similar states.
From this network, node $v$ computes the degree of each recurrence node,
\begin{equation*}
k_i^{(v)}(t)=\sum_{j=1}^{N}R_{ij}^{(v)}(t),
\end{equation*}
and forms the empirical degree distribution
\begin{equation*}
p_v(k;t).
\end{equation*}
This degree distribution is used as a compressed signature of the recent inventory dynamics. 
The key reason is that it preserves useful dynamical information while being much lighter and more privacy-preserving than sharing raw inventory data. 
Therefore, each node communicates only $P_v(k;t)$, not the original time series.
\paragraph{\textbf{Step 2: Privacy Preserving Sharing and Collaboration}}

After obtaining its local recurrence-network summary, node $v$ exchanges
$p_v(k;t)$ with same-tier peers. For each peer $u\in S(v)$, the similarity
between their local recurrence structures is measured using the
Jensen--Shannon divergence between their degree distributions.

First, we define the average distribution
\begin{equation*}
m_{vu}(k;t)
=
\frac{1}{2}
\left[
p_v(k;t)+p_u(k;t)
\right].
\end{equation*}
The Jensen--Shannon divergence is then given by
\begin{equation*}
D_{\mathrm{JS}}\!\left(p_v,p_u\right)
=
\frac{1}{2}
D_{\mathrm{KL}}\!\left(p_v\|m_{vu}\right)
+
\frac{1}{2}
D_{\mathrm{KL}}\!\left(p_u\|m_{vu}\right),
\end{equation*}
where
\begin{equation*}
D_{\mathrm{KL}}\!\left(p_v\|m_{vu}\right)
=
\sum_k
p_v(k;t)
\log
\frac{p_v(k;t)}{m_{vu}(k;t)},
\end{equation*}
and $D_{\mathrm{KL}}\!\left(p_u\|m_{vu}\right)$ is defined analogously.

Since a smaller Jensen--Shannon divergence indicates greater similarity,
we define the pairwise similarity index as
\begin{equation*}
I_{vu}(t)
=
1-
\frac{
D_{\mathrm{JS}}\!\left(p_v,p_u\right)
}{
\log 2
}.
\end{equation*}
Thus, $I_{vu}(t)\in[0,1]$, where larger values indicate greater similarity
between the local recurrence structures of nodes $v$ and $u$. In the
context of supply chains, an increase in $I_{vu}(t)$ suggests that the two
firms may be responding to the same upstream disturbance.

To obtain a single monitoring statistic at node $v$, we average over
same-tier peers:
\begin{equation*}
\bar I_v(t)
=
\frac{1}{|S(v)|}
\sum_{u\in S(v)}
I_{vu}(t).
\end{equation*}
An increase in $\bar I_v(t)$ indicates that node $v$ is becoming
dynamically more similar to its same-tier peers. This statistic is used
as the early-warning signal in the proposed framework.
\paragraph{\textbf{Step 3: Change Point Detection}}
The sequence $\{\bar I_v(t)\}$ is monitored over time to detect abrupt upward shifts.
Rather than running localisation continuously, node $v$ first checks whether the average similarity has undergone a statistically significant increase. 
In the current implementation, this is conducted using PELT with CROPS to obtain robust changepoint estimates \cite{Killick2012PELT,Haynes2017CROPS}.
Let $\tau_v^\ast$ denote the first significant upward changepoint. 
If no changepoint is detected, node $v$ remains in normal monitoring mode. 
If a changepoint is detected, node $v$ enters localisation mode.
This event-triggered mechanism suppresses unnecessary communication during stable operation while preserving sensitivity to disruption onset.
\paragraph{\textbf{Step 4: Local Extraction of Affected Clusters}}
After triggering, node $v$ builds a local similarity graph $H_v(t)$ over same-tier peers. 
Each vertex represents a peer node, and edges are weighted by the pairwise similarity index \(I_{u_1u_2}(t)\). Only similarities above a threshold $\eta$ are retained, so that the graph focuses on the strongest dynamic couplings. 
Let $A_v$ denote the adjacency matrix of this graph:
\begin{equation*}
(A_v)_{u_1u_2}=
\begin{cases}
I_{u_1u_2}(t), & I_{u_1u_2}(t)>\eta,\\
0, & \text{otherwise}.
\end{cases}
\end{equation*}
The corresponding graph Laplacian is
\begin{equation*}
L_v=D_v-A_v,
\end{equation*}
where $D_v$ is the degree matrix. Node $v$ then computes the Fiedler vector of $L_v$ and uses its sign structure to partition the graph into correlated groups \cite{Luxburg2007Spectral}. 
The affected set $C_v(t)$ is defined as the group containing node $v$ itself.
Intuitively, this step identifies which same-tier firms are evolving coherently after the warning signal appears. 
If a upstream edge has been disrupted, downstream firms depending on that edge tend to exhibit similar abnormal inventory dynamics, and thus are likely to be grouped into the same affected cluster.
\paragraph{\textbf{Step 5: Hierarchical Edge localisation}}
The affected set $C_v(t)$ is reported to upstream parent nodes. 
For an upstream node $p$, child reports are aggregated as
\begin{equation*}
C_p^{\mathrm{down}}(t)=\bigcup_{v\in \operatorname{ch}(p)} C_v(t).
\end{equation*}
If node $p$ also computes its own affected set $C_p(t)$, it compares the downstream evidence with its local evidence using the overlap score
\begin{equation*}
\mathrm{ov}_p(t)=
\frac{|C_p^{\mathrm{down}}(t)\cap C_p(t)|}
{|C_p^{\mathrm{down}}(t)\cup C_p(t)|}.
\end{equation*}
This upper-tier step is intentionally different from the direct same-tier clustering used at the downstream tier.
%
The reason is that disruption effects are typically first observed near directly affected downstream firms and may attenuate as they propagate across additional tiers, as inventory buffers and local resilience mechanisms can mitigate cascading effects \cite{lucker2019inventory,colon2020fragmentation,namdar2024cascading,hao2026continualevo,hao2026selfdis}.
Therefore, the similarity structure observed at higher tiers is usually weaker and more sensitive to the threshold used for graph construction.
If the same direct clustering procedure were repeatedly applied at upper tiers, a small threshold could easily turn weak or noisy correlations into artificial clusters, leading to overfitting and unstable localisation.
Instead, the upstream node uses overlap-based comparison to verify whether the downstream abnormality is also reflected in its own local dynamics. 
A low overlap suggests that the downstream nodes are affected while node $p$ remains relatively normal, indicating that the disruption is likely located on one of the outgoing edges of $p$. 
A high overlap suggests that the disturbance has already propagated to node $p$, and the evidence should therefore be merged and propagated further upstream.
By repeating this comparison tier by tier, the framework progressively narrows the candidate set and localises the disrupted supplier--buyer edge. 
In this way, the framework uses direct clustering where the disruption signal is strongest and overlap-based evidence integration where the signal becomes weaker, which improves the robustness of hierarchical localisation.

%
\begin{algorithm}[t]
\caption{Node-level distributed disruption monitoring and localisation}
\label{alg:node_level}
\footnotesize
\begin{algorithmic}[1]
\Require Window length $L$, recurrence rate $\rho$, similarity threshold $\eta$, overlap threshold $\theta_{\mathrm{overlap}}$
\Require Node $v$ with local signal $\{x_v(t)\}$, same-tier peer set $\mathcal{S}(v)$, parent set $\operatorname{pa}(v)$
\State initialise monitoring sequence $\{\bar I_v(t)\}\gets \emptyset$
\State initialise local affected set $C_v(t)\gets \emptyset$
\State initialise candidate disrupted-edge set $\widehat{\mathcal D}_v(t)\gets \emptyset$

\For{each decision epoch $t=1,2,\dots,T$}
    \State Update sliding window $W_v(t)=\{x_v(t-L+1),\ldots,x_v(t)\}$
    
    \State Construct the local recurrence network from $W_v(t)$
    \State Compute the empirical degree-distribution summary $P_v(k;t)$
    
    \State Exchange $P_v(k;t)$ with same-tier peers $u\in\mathcal{S}(v)$
    \For{each $u\in\mathcal{S}(v)$}
        \State Compute pairwise similarity index $I_{vu}(t)$
    \EndFor
    
    \State Compute average same-tier similarity $\bar I_v(t)=\frac{1}{|\mathcal{S}(v)|}\sum_{u\in\mathcal{S}(v)} I_{vu}(t)$
    \State Append $\bar I_v(t)$ to the monitoring sequence
    
    \State Run changepoint detection on $\{\bar I_v(\tau)\}_{\tau\le t}$
    
    \If{a significant upward changepoint is detected}
        \State Build the local similarity graph $H_v(t)$ from peer pairwise similarity values using threshold $\eta$
        \State Compute the graph Laplacian $L_v=D_v-A_v$
        \State Extract the affected set $C_v(t)$ by spectral partition of $L_v$
        \State Send $C_v(t)$ to all parent nodes in $\operatorname{pa}(v)$
    \EndIf
    
    \State Receive child reports from downstream nodes, if any
    \If{$v$ has received at least one child report}
        \State Aggregate downstream evidence $C_v^{\mathrm{down}}(t)=\bigcup_{w\in \operatorname{ch}(v)} C_w(t)$
        
        \If{$C_v(t)\neq\emptyset$ and $C_v^{\mathrm{down}}(t)\neq\emptyset$}
            \State Compute overlap score $\mathrm{ov}_v(t)=\frac{|C_v^{\mathrm{down}}(t)\cap C_v(t)|}{|C_v^{\mathrm{down}}(t)\cup C_v(t)|}$
            
            \If{$\mathrm{ov}_v(t)<\theta_{\mathrm{overlap}}$}
                \State Mark outgoing child edge(s) of node $v$ as disruption candidates
                \State Update $\widehat{\mathcal D}_v(t)$
            \Else
                \State Merge local and downstream evidence $C_v(t)\gets C_v(t)\cup C_v^{\mathrm{down}}(t)$
                \State Forward the merged affected set to upstream parent nodes
            \EndIf
        \EndIf
    \EndIf
\EndFor

\State \Return Local alert sequence, affected sets $\{C_v(t)\}$, and candidate disrupted-edge set $\widehat{\mathcal D}_v(t)$
\end{algorithmic}
\end{algorithm}
\paragraph{\textbf{Summary}}
The methodology combines local nonlinear-dynamics encoding, privacy-preserving peer comparison, changepoint detection, and hierarchical evidence aggregation into a single distributed monitoring framework. 
By communicating only recurrence-network summaries and triggering localisation only when coordination patterns change, the approach balances early detection, edge-level localisation, communication efficiency, and firm-level privacy.

\begin{algorithm}[t]
\caption{System-level distributed monitoring and edge localisation}
\label{alg:system_level}
\footnotesize
\begin{algorithmic}[1]
\Require Supply network $G=(V,E)$, monitoring horizon $T$
\Require Node-level policy in Algorithm~\ref{alg:node_level}
\State initialise node agents for all $v\in V$
\State initialise global candidate disrupted-edge set $\widehat{\mathcal D}(t)\gets \emptyset$
\State initialise confirmed disrupted-edge set $\widehat{\mathcal D}^{\,\mathrm{conf}}\gets \emptyset$

\For{each decision epoch $t=1,2,\dots,T$}
    \For{each node $v\in V$}
        \State Run Algorithm~\ref{alg:node_level} for node $v$ at time $t$
    \EndFor
    
    \State Collect all triggered alerts, local affected sets, and candidate edge reports
    \State Aggregate upstream messages tier by tier
    
    \For{each upstream node $p$ receiving child reports}
        \State Form aggregated downstream evidence $C_p^{\mathrm{down}}(t)=\bigcup_{v\in \operatorname{ch}(p)} C_v(t)$
        
        \If{$C_p(t)\neq\emptyset$ and $C_p^{\mathrm{down}}(t)\neq\emptyset$}
            \State Compute overlap score $\mathrm{ov}_p(t)=\frac{|C_p^{\mathrm{down}}(t)\cap C_p(t)|}{|C_p^{\mathrm{down}}(t)\cup C_p(t)|}$
            
            \If{$\mathrm{ov}_p(t)<\theta_{\mathrm{overlap}}$}
                \State Add outgoing child edge(s) of node $p$ to $\widehat{\mathcal D}(t)$
            \Else
                \State Forward merged evidence to the next upstream tier
            \EndIf
        \EndIf
    \EndFor
    
    \State De-duplicate candidates in $\widehat{\mathcal D}(t)$
    \State Update the system alert registry
    
    \If{the persistence criterion is satisfied}
        \State Emit confirmed disruption alert(s)
        \State Update $\widehat{\mathcal D}^{\,\mathrm{conf}}$
    \EndIf
\EndFor

\State \Return Confirmed disrupted-edge set $\widehat{\mathcal D}^{\,\mathrm{conf}}$, detection timestamps, and communication statistics
\end{algorithmic}
\end{algorithm}

\section{Experimental Setup}\label{sec:setup}
In this section, we describe the experimental setup used to evaluate the proposed decentralised early-warning framework and is organised into three components. 
Subsection \ref{sub: simulation} describes the agent-based simulation framework,
We develop an open-source agent-based simulator that models the structural dependencies and operational dynamics of multi-tier supply chains, use it to generate firm-level inventory trajectories under normal and disrupted conditions, and construct a dynamic inventory dataset for benchmarking decentralised disruption detection and edge localisation. Subsection \ref{sub:Sensitivity Study Design} examines the robustness of the proposed framework under varying disruption severity, supply-chain topology, and distance from the disrupted edge. Subsection \ref{sub: metrics} introduces the evaluation metrics used to assess edge-localisation performance, including detection rate, detection delay, precision, recall, and accuracy.

\subsection{Simulation Methodology}
\label{sub: simulation}
To evaluate the proposed decentralised disruption early-warning and localisation framework, we construct an agent-based simulation of a multi-tier supply chain in which firms interact through explicit supplier--buyer relationships induced by product input dependencies. 
The simulator generates firm-level inventory trajectories under both normal and disrupted operating conditions, and these trajectories are later used as the sole temporal input to the detection algorithm.
\begin{figure}[!b]
\centering
\includegraphics[width=\columnwidth]{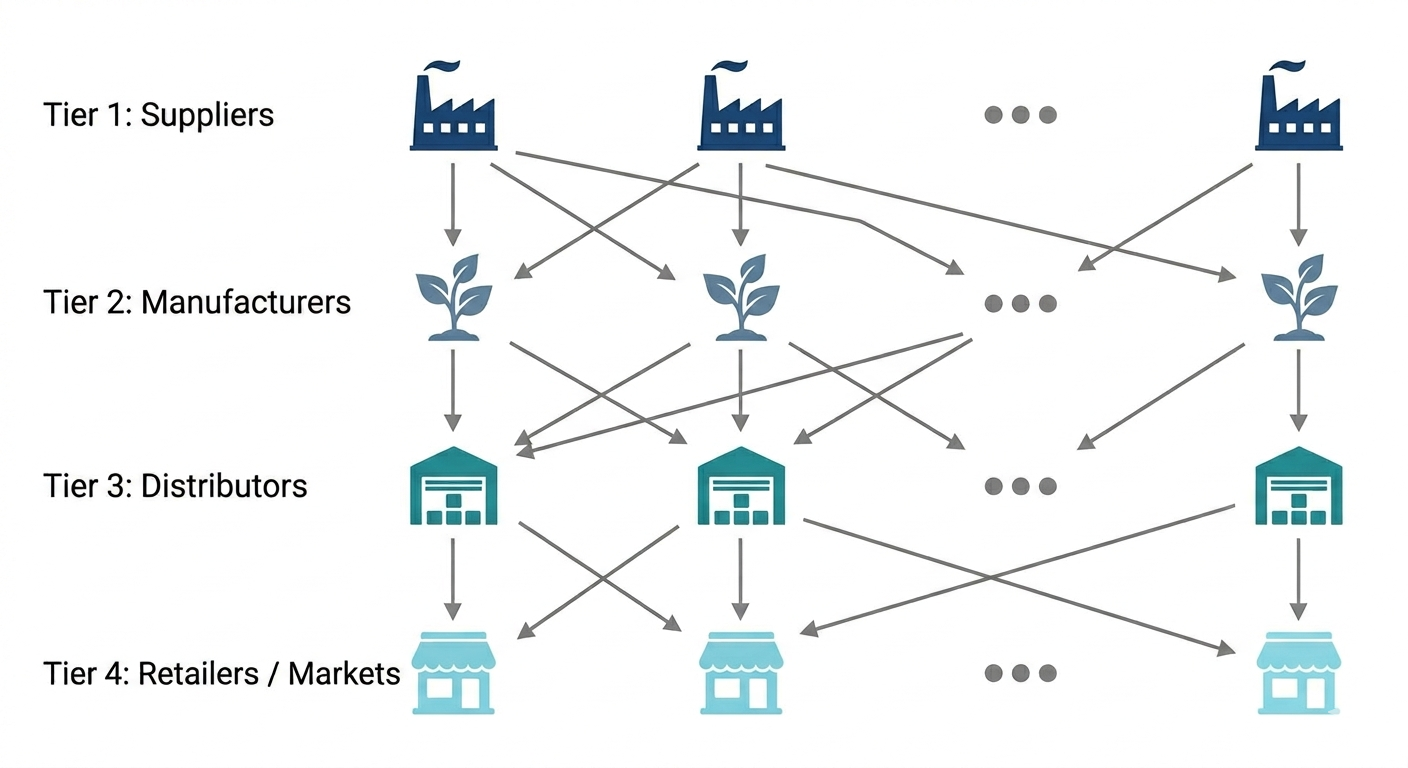}
\caption{Supply-chain graph used in the experimental setup. The graph illustrates the multi-tier supplier--buyer structure that underlies the simulation environment and disruption-propagation experiments.}
\label{fig:experimental_supply_chain_graph}
\end{figure}
Let $\mathcal{F}=\{f_1,\dots,f_{N_f}\}$ denote the set of firms and $\mathcal{P}=\{p_1,\dots,p_{N_p}\}$ denote the set of products. 
The supply chain is organised as a tiered directed acyclic graph. 
At the product level, we define a dependency graph
\begin{equation*}
G_P=(\mathcal{P},\mathcal{E}_P),
\end{equation*}
where an edge $(p_i,p_j)\in\mathcal{E}_P$ indicates that product $p_i$ is an input to product $p_j$.
Each such edge is associated with a units coefficient $u_{ij}>0$, which specifies how many units of $p_i$ are required to produce one unit of $p_j$. 
These coefficients are assembled into a product-input matrix
\begin{equation*}
\mathbf{M}\in\mathbb{R}^{N_p\times N_p},
\qquad
M_{j,i}=u_{ij},
\end{equation*}
where row $j$ encodes the input recipe for product $p_j$. If row $j$ is all zeros, then $p_j$ is an exogenous product and does not require any upstream input for production.
At the firm level, each product is assigned to one or more firms capable of producing it. 
Let $\mathcal{P}(f)\subseteq\mathcal{P}$ denote the set of products supplied by firm $f$, and let $\mathcal{F}(p)\subseteq\mathcal{F}$ denote the set of firms capable of producing product $p$. 
For each downstream firm $b$ and each required input product $p$, exactly one active upstream supplier $s$ is selected from the previous tier.
This defines a supplier map
\begin{equation*}
\phi:(b,p)\mapsto s,
\end{equation*}
and induces a directed supplier--buyer graph
\[
\begin{aligned}
G_F &= (\mathcal{F},\mathcal{E}_F),\\
(s,b)\in\mathcal{E}_F &\iff \exists p \text{ such that } \phi(b,p)=s.
\end{aligned}
\]
By construction, $G_F$ is tiered and acyclic, so disruptions propagate in a controlled upstream-to-downstream manner.
An example of the tiered supplier--buyer network generated by the simulator is shown in Figure~\ref{fig:experimental_supply_chain_graph}.
The dynamic state of the simulator at time $t$ consists primarily of the firm-product inventory matrix
\begin{equation*}
\mathbf{I}(t)\in\mathbb{R}_{\ge 0}^{N_f\times N_p},
\end{equation*}
where $I_{f,p}(t)$ is the inventory held by firm $f$ for product $p$. 
In addition, the simulator maintains an order book and an exogenous supply schedule.
For exogenous products, the available supply to firm $f$ for product $p$ at time $t$ is denoted by
\begin{equation*}
S^{\text{exo}}_{f,p}(t).
\end{equation*}
Consumer demand is imposed only on final products, i.e., products with no outgoing dependency edges. 
For each firm-product pair $(f,p)$ corresponding to a consumer product, stochastic demand $D_{f,p}(t)$ is generated from a time-varying demand schedule. 
In the current implementation, demand evolves as a drifting stochastic process with pattern-specific modulation and is realised using a Poisson model:
\begin{equation*}
D_{f,p}(t)\sim \operatorname{Poisson}(\lambda_{f,p}(t)),
\end{equation*}
where $\lambda_{f,p}(t)$ captures both slow drift and temporal pattern effects such as weekday-dominant, weekend-dominant, or uniform demand.
At each discrete time step $t=1,\dots,T$, the simulation evolves as follows.
First, newly realised consumer demand is appended to the current order book. 
Second, firms are processed in a random order to avoid artifacts caused by a fixed update sequence.
For each firm $f$ and each supplied product $p\in\mathcal{P}(f)$, the simulator attempts to fulfil outstanding orders.
If $p$ is exogenous, fulfilment is limited by available exogenous supply. 
If an order requests quantity $q$, then it can be fulfilled only if
\begin{equation*}
q \le S^{\text{exo}}_{f,p}(t).
\end{equation*}
If fulfilled, exogenous supply is decremented accordingly.
If $p$ is endogenous, order fulfilment depends on the firm's current inventories of all required inputs.
Let $\mathbf{m}_p$ denote the input vector for product $p$, i.e., the row of $\mathbf{M}$ corresponding to product $p$. 
To fulfil an order of size $q$, the firm must satisfy
\begin{equation*}
\mathbf{I}_{f,:}(t) \ge q\,\mathbf{m}_p
\end{equation*}
elementwise. 
If this condition holds, the necessary inputs are consumed:
\begin{equation*}
\mathbf{I}_{f,:}(t)\leftarrow \mathbf{I}_{f,:}(t)-q\,\mathbf{m}_p.
\end{equation*}
The completed shipment is then recorded as a transaction from the supplier to the buyer.
After all firms have processed their eligible orders, completed shipments are added to the buyers' inventories. 
Thus, if $X_{s,b,p}(t)$ denotes the quantity of product $p$ shipped from firm $s$ to firm $b$ at time $t$, then the buyer-side inventory update can be written as
\begin{equation*}
I_{b,p}(t+1)=I_{b,p}(t)+\sum_{s} X_{s,b,p}(t),
\end{equation*}
after accounting for any production-side consumption already applied during fulfilment.
Once current shipments have been executed, firms compute the remaining inputs required to satisfy unfinished downstream orders.
Let $R_{f,p}(t)$ denote the residual required quantity of product $p$ at firm $f$ after considering current inventories and pending receipts. 
New replenishment orders are then placed upstream. 
Under the default deterministic sourcing policy, firm $f$ orders product $p$ from its assigned supplier $\phi(f,p)$. Therefore, the order flow is constrained by the supplier map and the product dependency structure.
A disruption is injected by selecting a supplier--buyer edge $(s^\star,b^\star)\in\mathcal{E}_F$ and a disruption onset time $t_0$.
The disruption is implemented by disabling all transactions from $s^\star$ to $b^\star$ for $t\ge t_0$. 
Formally,
\begin{equation*}
X_{s^\star,b^\star,p}(t)=0,
\qquad \forall p,\; t\ge t_0.
\end{equation*}
All other transactions proceed normally. 
This intervention creates a known edge-level failure with exact ground-truth location and onset time.
In the reported experiments, exogenous shocks are disabled, so the only abnormality in the system arises from the injected supplier--buyer disruption itself.
Although the simulator tracks inventories at the firm-product level, the detection algorithm operates on firm-level total inventory.
Therefore, after simulation, each firm $f$ is assigned an aggregated inventory trajectory
\begin{equation*}
x_f(t)=\sum_{p\in\mathcal{P}} I_{f,p}(t),
\qquad t=0,\dots,T.
\end{equation*}
This aggregation produces a one-dimensional time series per firm, which serves as the input to the recurrence-network-based detection stage. 
Hence, the simulator provides the detection tier with the set
\begin{equation*}
\{x_f(t)\}_{f\in\mathcal{F},\, t=0}^{T},
\end{equation*}
together with the supplier--buyer network $G_F$ and the known disrupted edge $(s^\star,b^\star)$.
Overall, this simulation framework provides a controlled but operationally meaningful benchmark.
It preserves the interaction between product dependency constraints, supplier--buyer structure, stochastic demand realisation, and inventory evolution, while enabling rigorous evaluation through known edge-level disruption injection.
As a result, it is well suited for validating both early-warning performance and disruption localisation accuracy under limited shared information.
\paragraph{\textbf{Dataset}}
Based on the simulation framework, we construct a dynamic inventory dataset for evaluating decentralised disruption early warning and edge localisation. 
Each sample corresponds to one simulated four-tier supply chain instance together with its firm-level inventory sequences and disruption labels.
Specifically, the firm set is organised as
\[
\mathcal{F}=\mathcal{F}^{(1)}\cup \mathcal{F}^{(2)}\cup \mathcal{F}^{(3)}\cup \mathcal{F}^{(4)},
\]
where each tier contains a random number of firms:
\begin{equation*}
N_\ell \sim \mathrm{Unif}\{10,11,\dots,20\}, \qquad \ell=1,2,3,4.
\end{equation*}
Thus, each supply chain instance contains between 40 and 80 firms in total.
Supplier--buyer relationships between adjacent tiers are generated according to the product dependency structure, resulting in a tiered directed acyclic graph
\[
G_F=(\mathcal{F},\mathcal{E}_F).
\]
For each firm $f\in\mathcal{F}$, the simulator records a one-dimensional inventory sequence of length $T=100$:
\begin{equation*}
x_f(t), \qquad t=1,\dots,100.
\end{equation*}
Here, $x_f(t)$ denotes the aggregated inventory of firm $f$ at time $t$. 
Therefore, each sample can be represented as
\begin{equation*}
\mathcal{S}=\bigl(G_F,\mathcal{X},\mathcal{D}^\star,t_0\bigr),
\end{equation*}
where
\[
\mathcal{X}=\{x_f(t)\}_{f\in\mathcal{F},\,t=1}^{100}
\]
is the set of firm-level inventory sequences, $\mathcal{D}^\star$ is the ground-truth disrupted-edge set, and $t_0$ is the disruption onset time for disrupted cases.
For disrupted samples, one supplier--buyer edge
\[
e^\star=(s^\star,b^\star)\in\mathcal{E}_F
\]
is selected and disabled from time $t_0$ onward, so that
\[
\mathcal{D}^\star=\{e^\star\}.
\]
For non-disrupted samples, no disruption is injected and $\mathcal{D}^\star=\emptyset$.
Overall, the dataset provides graph-structured supply chain instances with variable firm populations and length-100 firm-level inventory sequences, making it suitable for evaluating both early warning and edge localisation under limited shared information.
\subsection{Sensitivity Study Design}
\label{sub:Sensitivity Study Design}
To further examine the robustness of the proposed framework, we extend the experiments along three dimensions: disruption severity, supply-chain topology, and the distance between the downstream firm and the disrupted edge.
\paragraph{Disruption Severity}
In the main experiments, a disrupted edge is fully disabled after the disruption onset time. 
To model partial degradation, we introduce a severity parameter $\alpha\in[0,1]$ and define the post-disruption transaction quantity on the disrupted edge as
\begin{equation}
X_{s^\star,b^\star,p}^{\mathrm{post}}(t)
=
(1-\alpha)\,X_{s^\star,b^\star,p}^{\mathrm{normal}}(t),
\qquad t\ge t_0,
\end{equation}
where $\alpha=1$ corresponds to complete disruption and smaller values correspond to partial loss of supply capacity.
In the experiments, we evaluate
\begin{equation}
\alpha\in\{0.25,0.5,0.75,1.0\}.
\end{equation}
\paragraph{Supply-Chain Topology}
We compare three structural settings.
The first is a \emph{strictly hierarchical topology}, in which edges are allowed only between adjacent tiers, i.e., from tier $\ell$ to tier $\ell+1$.
The second is a \emph{cross-tier topology}, in which a subset of firms can form direct supply links that skip intermediate tiers.
Formally, let
\begin{equation}
E_{\mathrm{skip}}=\{(u,v)\in E:\ell(v)-\ell(u)\ge 2\}
\end{equation}
denote the set of skip-tier edges.
In the strictly hierarchical setting, $E_{\mathrm{skip}}=\emptyset$, whereas in the cross-tier setting, skip connections are introduced with probability $\rho=0.25$ and maximum span $d_{\max}=2$.
The third is a \emph{scale-free topology}, in which the in-degree or out-degree distribution follows a power-law pattern, with a small number of hub firms connected to many partners and most firms connected to only a few.
\paragraph{Distance from the Disrupted Edge}

We further test whether localisation performance depends on how far the disrupted edge is from the monitored downstream firm.
Under the strictly hierarchical topology and complete disruption, we inject disruptions at three upstream positions.
Distance 0 denotes an edge directly connected to the downstream firm, whereas distances 1 and 2 denote edges one and two tiers upstream.
This experiment examines whether the framework can capture both directly connected disruptions and propagated upstream effects.

\subsection{ Evaluation Metrics}
\label{sub: metrics}
The main outputs of the simulation-detection pipeline are a detection log, a candidate set of disrupted edges, and the first detection time of the true disrupted edge.
Using the known ground truth, performance is evaluated through standard edge-level metrics.
Let $\mathcal{D}^\star$ denote the true disrupted-edge set and let $\widehat{\mathcal{D}}$ denote the candidate set returned by the method.
Precision and recall are defined as
\[
\mathrm{Precision}=\frac{|\widehat{\mathcal{D}}\cap \mathcal{D}^\star|}{|\widehat{\mathcal{D}}|},
\qquad
\mathrm{Recall}=\frac{|\widehat{\mathcal{D}}\cap \mathcal{D}^\star|}{|\mathcal{D}^\star|}.
\]
For single-edge disruption experiments, the exact detection rate is the fraction of trials in which the injected edge appears in the candidate set. For multi-edge disruption experiments, the partial detection rate is the fraction of trials in which at least one true disrupted edge is identified.
The detection delay is defined as
\begin{equation*}
\Delta t = t_{\mathrm{detect}} - t_0,
\end{equation*}
where $t_{\mathrm{detect}}$ is the first time at which the true disrupted edge is identified by the method.

\section{Experimental Results}\label{sec:results}
\begin{table*}[t]
\centering
\small
\setlength{\tabcolsep}{5pt}
\renewcommand{\arraystretch}{1.1}
\caption{Main edge-localisation performance under increasing disruption complexity. }
\label{tab:main_results}
\begin{tabular}{@{}lcccc@{}}
\toprule
\shortstack[c]{Disrupted Edges\\ per Trial} & Accuracy  & Recall & \shortstack[c]{Exact / All-Edges\\ Detection Rate} & \shortstack[c]{Partial\\ Detection Rate} \\
\midrule
1 & $0.73 \pm 0.06$  & $0.80 \pm 0.14$ & $0.80 \pm 0.14$ & -- \\
2 & $0.74 \pm 0.08$  & $0.52 \pm 0.11$ & $0.24 \pm 0.12$ & $0.80 \pm 0.13$ \\
3 & $0.64 \pm 0.08$  & $0.40 \pm 0.07$ & $0.05 \pm 0.02$ & $0.92 \pm 0.07$  \\
\bottomrule
\end{tabular}
\end{table*}
The experiments were conducted on a four-tier synthetic supply chain generated by the agent-based simulator. 
In each trial, one or multiple supplier-buyer edges were randomly selected and disrupted at time step 40. 
The proposed decentralised detection algorithm was then applied to firm-level inventory time series to identify candidate disrupted edges. 
The results are presented in two parts. 
First, we compare the proposed method across three disruption settings and against a baseline method in Section~\ref{main result}. 
Second, we evaluate its robustness under varying disruption severity, supply-chain topology, and distance from the disrupted edge in Section~\ref{robustness study}.
\begin{table*}[t]
\centering
\small
\setlength{\tabcolsep}{5pt}
\renewcommand{\arraystretch}{1.1}
\caption{Detection count and mean delay comparison between the proposed method and a local inventory baseline.}
\label{tab:baseline_delay}
\begin{tabular}{@{}lccc@{}}
\toprule
Method & Detection rate & Mean Delay & Accuracy \\
\midrule
Proposed method & $0.80 \pm 0.14$ & $11.00 \pm 9.73$ & $0.73 \pm 0.06$ \\
\shortstack[l]{Local inventory baseline} & $0.21 \pm 0.12$ & $1.00  \pm 1.93$ & $0.47 \pm 0.02$ \\
\bottomrule
\end{tabular}
\end{table*}

\subsection{Main Result}~\label{main result}

Table~\ref{tab:main_results} reports performance under single-edge, two-edge, and three-edge disruption settings.
In the single-edge case, the proposed method achieves an exact detection rate of $0.80$, showing that it can recover the true disrupted supplier--buyer link under decentralised information constraints. 
As the number of simultaneous disruptions increases, exact recovery becomes harder, but the method still identifies at least one true disrupted edge in most trials: the partial detection rate is $0.80$ for two-edge disruptions and $0.92$ for three-edge disruptions. 
These results indicate that the framework is not limited to isolated failures and can retain useful localisation capability under more complex disruption patterns.
Table~\ref{tab:baseline_delay} compares the proposed method with a local inventory-threshold baseline, which raises an alert when the downstream buyer's inventory falls below the 5th percentile of its pre-disruption distribution. 
This baseline detects only $21$ out of $100$ single-edge disruptions, compared with $80$ detections by the proposed method. 
The result suggests that edge disruptions are often not visible as immediate local inventory depletion, because buffer inventory and demand variability can mask early effects. 
By using inter-firm dynamical similarity and upstream evidence propagation, the proposed method provides substantially stronger edge-localisation performance than local threshold monitoring.
\subsection{Robustness Study}~\label{robustness study}
We further investigate how disruption severity, supply-chain topology, and the distance between the downstream firm and the disrupted edge affect edge-localisation performance.
In particular, we report detection rate, detection delay, recall, and accuracy where applicable.
These metrics together indicate whether the true disrupted edge can be recovered, how rapidly the framework identifies it after disruption onset, and how performance changes as the disruption occurs farther from the downstream firm.
\subsubsection{Effect of Disruption Severity}
We first evaluate the impact of partial edge degradation under the strictly hierarchical topology. 
The disrupted edge is scaled according to the severity parameter $\alpha\in\{0.25,0.5,0.75,1.0\}$, where $\alpha=1.0$ corresponds to complete disruption. 
The results are summarised in the left panel of Table~\ref{tab:severity_delay}.
\begin{table*}[t]
\centering
\small
\setlength{\tabcolsep}{4pt}
\renewcommand{\arraystretch}{1.1}
\caption{Robustness analysis under varying disruption severity and supply-chain topology. The left panel reports detection performance under partial edge degradation in the strictly hierarchical topology, while the right panel compares detection delay across strictly hierarchical, cross-tier  and scale-free topologies under complete disruption.}
\label{tab:severity_delay}
\label{tab:topology_delay}
\begin{minipage}[t]{0.49\textwidth}
\centering
\textbf{(a) Disruption severity}\\[0.35em]
\begin{tabular}{@{}cccc@{}}
\toprule
Severity $\alpha$ & \shortstack[c]{Detection\\ Rate} & \shortstack[c]{Mean\\ Delay} & Accuracy  \\
\midrule
0.25 & $0.43 \pm 0.24$ & $22.25 \pm 9.50$  & $0.86 \pm 0.07$  \\
0.50 & $0.45 \pm 0.18$ & $24.00 \pm 6.85$  & $0.84 \pm 0.08$  \\
0.75 & $0.65 \pm 0.07$ & $13.33 \pm 11.75$ & $0.82 \pm 0.07$\\
1.00 & $0.80 \pm 0.14$ & $11.00 \pm 9.73$  & $0.73 \pm 0.06$ \\
\bottomrule
\end{tabular}
\end{minipage}\hfill
\begin{minipage}[t]{0.47\textwidth}
\centering
\textbf{(b) Supply-chain topology}\\[0.35em]
\begin{tabular}{@{}lccc@{}}
\toprule
Topology & \shortstack[c]{Detection\\ Rate} & \shortstack[c]{Mean\\ Delay}  & Accuracy \\
\midrule
Strictly hierarchical & $0.80 \pm 0.14$  & $11.00 \pm 9.73$ & $0.73 \pm 0.06$ \\
Cross-tier & $0.40 \pm 0.17$ & $32.00 \pm 10.76$ & $0.67 \pm 0.06$ \\
Scale-free & $0.82 \pm 0.09$ & $21.46 \pm 10.92$ & $0.76 \pm 0.08$ \\
\bottomrule
\end{tabular}
\end{minipage}
\end{table*}
Overall, stronger disruptions tend to produce shorter localisation delays. 
In particular, the mean delay decreases from 22.25 time steps at $\alpha=0.25$ to 11.00 time steps at $\alpha=1.0$. 
This trend is consistent with the intuition that stronger disruptions induce larger deviations in downstream inventory dynamics, which in turn generate more pronounced same-tier synchronisation and earlier changepoint triggering.
The delay pattern suggests that partial disruptions are generally more difficult to distinguish from normal operational fluctuations than complete edge failures.
\subsubsection{Effect of Supply-Chain Topology}
We next examine the effect of network structure under complete disruption ($\alpha=1.0$). 
Specifically, we compare the strictly hierarchical topology with the cross-tier topology  and scale-free topology described above. 
The results are shown in the right panel of Table~\ref{tab:topology_delay}.
The results indicate that cross-tier connectivity makes edge localisation substantially more difficult in the current setting. 
Compared with the strictly hierarchical topology, the cross-tier topology yields both a lower detection rate and a much larger localisation delay. 
%
Specifically, the mean delay increases from 11.00 to 32.00 time steps. 
A plausible explanation is that cross-tier links introduce additional propagation paths and weaken the regularity of tier-wise anomaly transmission. 
Previous studies have highlighted the complex and heterogeneous structure of real-world supply networks, including the presence of scale-free characteristics \cite{brintrup2018supply,hearnshaw2013complex,choi2001supply,wiedmer2021structural}. 
Therefore, evaluating the proposed framework under scale-free topology is particularly relevant to assessing its practical applicability to real-world supply chains.
The scale-free topology shows a different pattern.
Although its mean delay increases from 11.00 to 21.46 time steps compared with the strictly hierarchical topology, it remains lower than that of the cross-tier topology.
This suggests that hub firms in scale-free networks can concentrate disruption effects and make the propagation signal more detectable.


%
However, the highly uneven degree distribution also makes same-tier comparison less uniform: firms connected to hubs may exhibit strong correlated changes, whereas peripheral firms with few links may show weaker or delayed responses.
Therefore, scale-free networks can provide clearer disruption signals around hubs, but their heterogeneous connectivity still makes localisation less timely than in regular strictly hierarchical networks.
\subsubsection{Effect of Distance from the Disrupted Edge}
\begin{table*}[t]
\centering
\small
\setlength{\tabcolsep}{5pt}
\renewcommand{\arraystretch}{1.1}
\caption{Detection performance under different distances between the downstream firm and the disrupted edge.}
\label{tab:downstream_distance_detection}
\begin{tabular}{@{}ccccc@{}}
\toprule
Distance & \shortstack[c]{Disrupted\\ Edge tier} & \shortstack[c]{Detection\\ Rate} & \shortstack[c]{Mean\\ Delay}  & Accuracy \\
\midrule
0 & L2 $\rightarrow$ L3 & $0.70 \pm 0.20$ & $18.29 \pm 8.02$  & $0.79 \pm 0.07$ \\
1 & L1 $\rightarrow$ L2 & $0.70 \pm 0.17$ & $13.73 \pm 6.44$  & $0.64 \pm 0.09$ \\
2 & L0 $\rightarrow$ L1 & $0.80 \pm 0.14$ & $19.62 \pm 13.07$  & $0.73 \pm 0.08$\\
\bottomrule
\end{tabular}
\end{table*}
We evaluate whether detection performance changes with the distance between the downstream retailer and the disrupted edge. 
Distance 0 denotes a disruption on the edge directly connected to the retailer, while distances 1 and 2 denote disruptions one and two tiers upstream.
The exact detection rate is 0.70 for distance 0 and distance 1, and increases to 0.80 for distance 2. The mean detection delay is lowest 13.73 time steps at distance 1 , compared with 18.29 and 19.62 for distances 0 and 2.
Overall, the effect of distance is not strictly monotonic. The method performs best for farther upstream disruptions, suggesting that it detects propagated network-level changes more effectively than purely local disruptions.

Taken together, the robustness results show that the proposed framework is affected by disruption strength, network topology, and the relative position of the disrupted edge.
Stronger disruptions are localised more quickly because they generate clearer temporal-topological changes in downstream inventory trajectories.
Topology also plays an important role: cross-tier links introduce additional propagation paths and make same-tier synchronisation less coherent, leading to longer localisation delays.
Scale-free networks show an intermediate pattern, where hub firms can concentrate disruption signals, but heterogeneous connectivity still makes localisation less timely than in regular strictly hierarchical networks.
The distance analysis further shows that performance is not strictly monotonic with the distance between the monitored downstream firm and the disrupted edge, suggesting that the framework can capture both local disruptions and propagated network-level effects.
Overall, the method is most effective in regular tiered supply-chain structures, while more path-aware or hub-aware evidence fusion may be needed for networks with skip-tier links, scale-free connectivity, or complex propagation patterns.
\section{Conclusion}\label{sec:conclusion}

This paper introduced a distributed early-warning framework for detecting and localising supplier--buyer edge disruptions under limited information sharing. The proposed method combines local recurrence-network summaries, same-tier similarity monitoring, changepoint detection, affected-cluster extraction, and hierarchical upstream tracing, allowing firms to collaborate without exchanging raw operational trajectories. Experiments on an agent-based multi-tier supply-chain simulator show that the framework can localise disrupted edges under both single-edge and multi-edge settings and outperform a local inventory-threshold baseline.

The robustness analysis further shows how disruption severity, supply-chain topology, and the distance from the disrupted edge affect detection and localisation. Higher disruption severity generally leads to shorter localisation delays because it produces more pronounced changes in downstream inventory dynamics. Network topology also plays an important role: cross-tier connections introduce additional propagation paths, reducing detection performance and increasing localisation delays, whereas the framework maintains a high detection rate under scale-free topology despite its heterogeneous connectivity. Finally, detection rates and localisation delays do not vary monotonically with the distance from the disrupted edge, indicating that greater propagation distance does not necessarily lead to poorer detection performance in supply chains. Overall, these results demonstrate the potential of privacy-preserving temporal-topological monitoring for distributed disruption detection and edge-level localisation.

\subsection{Managerial Implications}

As organisations increasingly adopt digital supply chain twins and agentic AI frameworks, there is an opportunity to integrate privacy-preserving early-warning mechanisms into supply-chain monitoring systems. The proposed framework enables firms to detect and localise disruptions using distributed operational data without sharing raw information, reducing the need for centralised data collection. This provides a potential foundation for collaborative disruption monitoring across supply networks while preserving firm-level control over operational data.

\subsection{Limitations and Future Work}

The current study has several limitations. First, the framework is evaluated using synthetic data generated by an agent-based simulator, and its performance therefore remains to be validated using real-world operational data. Second, the current formulation assumes an approximately acyclic supply-chain structure, while real-world networks may contain cyclic, partially observed, and more irregular relationships.

A further practical limitation concerns the scale of deployment. Firms may favour small-scale adoption to minimise setup and maintenance costs, but limited participation may reduce the availability of cross-firm information and weaken early-warning performance. Broader participation could improve network coverage, but would require greater coordination and investment. This creates an adoption dilemma in which the effectiveness of distributed early-warning systems may depend on achieving sufficient participation across the supply network.

Future work will validate the framework using real or semi-synthetic supply-chain traces and extend it to partially observed and cyclic networks. More adaptive and topology-aware evidence-fusion mechanisms will also be explored to improve localisation in networks with cross-tier links, scale-free connectivity, and other irregular propagation structures. From a deployment perspective, future research should investigate participation incentives, large-scale coordination, and governance arrangements that can encourage wider adoption while preserving decentralised data ownership.


\bibliographystyle{plain}
\nocite{*}
\bibliography{references}

\appendix

\end{document}